\documentclass[numsec,webpdf,modern,large,namedate]{oup-authoring-template}

\onecolumn 

\graphicspath{{Fig/}}

\usepackage{comment}
\usepackage{amsmath, amssymb}

\theoremstyle{thmstyleone}%
\theoremstyle{thmstyletwo}%
\theoremstyle{thmstylethree}%

\makeatletter
\DeclareRobustCommand\vdots{%
  \mathpalette\@vdots{}%
}
\newcommand*{\@vdots}[2]{%
  \sbox0{$#1\cdotp\cdotp\cdotp\m@th$}%
  \sbox2{$#1.\m@th$}%
  \vbox{%
    \dimen@=\wd0 %
    \advance\dimen@ -3\ht2 %
    \kern.5\dimen@
    \dimen@=\wd2 %
    \advance\dimen@ -\ht2 %
    \dimen2=\wd0 %
    \advance\dimen2 -\dimen@
    \vbox to \dimen2{%
      \offinterlineskip
      \copy2 \vfill\copy2 \vfill\copy2 %
    }%
  }%
}
\DeclareRobustCommand\ddots{%
  \mathinner{%
    \mathpalette\@ddots{}%
    \mkern\thinmuskip
  }%
}
\newcommand*{\@ddots}[2]{%
  \sbox0{$#1\cdotp\cdotp\cdotp\m@th$}%
  \sbox2{$#1.\m@th$}%
  \vbox{%
    \dimen@=\wd0 %
    \advance\dimen@ -3\ht2 %
    \kern.5\dimen@
    \dimen@=\wd2 %
    \advance\dimen@ -\ht2 %
    \dimen2=\wd0 %
    \advance\dimen2 -\dimen@
    \vbox to \dimen2{%
      \offinterlineskip
      \hbox{$#1\mathpunct{.}\m@th$}%
      \vfill
      \hbox{$#1\mathpunct{\kern\wd2}\mathpunct{.}\m@th$}%
      \vfill
      \hbox{$#1\mathpunct{\kern\wd2}\mathpunct{\kern\wd2}\mathpunct{.}\m@th$}%
    }%
  }%
}
\makeatother

\usepackage{graphicx}
\newcommand\smallO{
  \mathchoice
    {{\scriptstyle\mathcal{O}}}
    {{\scriptstyle\mathcal{O}}}
    {{\scriptscriptstyle\mathcal{O}}}
    {\scalebox{.7}{$\scriptscriptstyle\mathcal{O}$}}
  }

\begin{document}

\journaltitle{Journal of the Royal Statistical Society, Series C: Applied Statistics}
\DOI{}
\copyrightyear{\the\year}
\pubyear{\the\year}
\access{Advance Access Publication Date:}
\appnotes{}

\firstpage{1}


\title[Personalised dynamic super learning]{Personalised dynamic super learning: an application in predicting hemodiafiltration convection volumes }

\author[1,$\ast$,\ORCID{0000-0002-0018-5899}]{Arthur Chatton}
\author[1,2,\ORCID{0000-0001-9587-7929}]{Mich\`ele Bally}
\author[3]{Ren\'ee L\'evesque}
\author[4]{Ivana Malenica}
\author[5,6,7,\ORCID{0000-0002-5981-8443}]{Robert W. Platt}
\author[1,7,8]{Mireille E. Schnitzer}

\authormark{Chatton et al.}

\address[1]{\orgdiv{Facult\'e de Pharmacie}, \orgname{Universit\'e de Montr\'eal}, \orgaddress{Montr\'eal, \state{QC}, \country{Canada}}}
\address[2]{\orgdiv{D\'epartement de Pharmacie et Centre de Recherche}, \orgname{Centre Hospitalier de l'Universit\'e de Montr\'eal}, \orgaddress{\street{Montr\'eal}, \state{QC}, \country{Canada}}}
\address[3]{\orgdiv{Service de N\'ephrologie, D\'epartement de M\'edecine}, \orgname{Universit\'e de Montr\'eal}, \orgaddress{\street{Montr\'eal}, \state{QC}, \country{Canada}}}
\address[4]{\orgdiv{Department of Statistics}, \orgname{Harvard University}, \orgaddress{\street{Cambridge}, \state{MA}, \country{USA}}}
\address[5]{\orgdiv{Department of Pediatrics}, \orgname{McGill University}, \orgaddress{\street{Montr\'eal}, \state{QC}, \country{Canada}}}
\address[6]{\orgdiv{Center for Clinical Epidemiology, Lady Davis Institute}, \orgname{Jewish General Hospital}, \orgaddress{\street{Montr\'eal}, \state{QC}, \country{Canada}}}
\address[7]{\orgdiv{Department of Epidemiology, Biostatistics, Occupational Health}, \orgname{McGill University}, \orgaddress{\street{Montr\'eal}, \state{QC}, \country{Canada}}}
\address[8]{\orgdiv{D\'epartement de M\'edecine Sociale et Pr\'eventive}, \orgname{Universit\'e de Montr\'eal}, \orgaddress{Montr\'eal, \state{QC}, \country{Canada}}}

\corresp[$\ast$]{\textbf{Correspondence}: Arthur Chatton. Faculté de pharmacie, Université de Montréal. Pavillon Jean-Coutu, 2940, chemin de Polytechnique, Montréal, QC, H3T-1J4, Canada. email:\href{email:arthur.chatton@umontreal.ca}{arthur.chatton@umontreal.ca}}

\received{Date}{0}{Year}
\revised{Date}{0}{Year}
\accepted{Date}{0}{Year}



\abstract{Obtaining continuously updated predictions is a major challenge for personalised medicine. Leveraging combinations of parametric regressions and machine learning algorithms, the personalised online super learner (POSL) can achieve such dynamic and personalised predictions. We adapt POSL to predict a repeated continuous outcome dynamically and propose a new way to validate such personalised or dynamic prediction models. We illustrate its performance by predicting the convection volume of patients undergoing hemodiafiltration. POSL outperformed its candidate learners with respect to median absolute error, calibration-in-the-large,  discrimination, and net benefit. We finally discuss the choices and challenges underlying the use of POSL.
}

\maketitle

\section{Introduction}\label{sec1}

Dynamic prediction models provide predicted outcome values that can be updated over time for an individual as new measurements become available \citep{Proust-Lima_Blanche_2016, Jenkins_2021}. Such models are crucial for providing updated predictions, giving users the most useful information at the current time. For instance, \citet{Fournier_2019} developed a dynamic prediction model incorporating repeated creatinine measurement to inform kidney transplant recipients of their prognosis and disease progression. \citet{Sabathe_2023} proposed a dynamic prediction model for guiding the early decision to switch from first- to second-line therapy in multiple sclerosis.  Outside of clinical fields, dynamic models have been used, for example, to predict geological sharp slope deformation \citep{Li_2021} and growth/decline of tourism demand in Australia \citep{Athanasopoulos_2009}.

Previous approaches to prediction were mainly based on parametric models, but there is a current trend towards so-called machine learning (ML) algorithms \citep{Breiman_2001, deHond_2022}. We stress that, for prediction, there is a false dichotomy between traditional “statistical” regression models and ML algorithms \citep{Finlayson_2023}. They are both statistical learning algorithms aiming to map some inputs to an output by optimising a loss function. Culturally, we typically refer to ML algorithms as methods that do not rely or rely to a lesser extent on parametric assumptions about the underlying data-generating process and have the potential to predict more accurately. Nevertheless, there is still no free-lunch method, even with ML \citep{Wolpert_1996}. Their improved flexibility is at the cost of an increased risk of over-fitting, needing more data to converge, and the final performance still depends on the particular data-generating process and the chosen hyperparameters \citep{Hand_2006}. 

Ensemble algorithms aim to circumvent the “best-algorithm choice” by combining a diverse and rich set of algorithms or “ candidate learners” into one final prediction \citep{Breiman_1996, Naimi_Balzer_2018}. For instance, the super learner \citep[SL,][]{vdl_Polley_Hubbard_2007} takes a weighted sum of the individual predictions from each incorporated learner (hereafter candidate learners) to produce a final prediction that is theoretically as accurate as the best-performing (with respect to a user-defined loss function) candidate learner \citep{Dudoit_vdL_2005}. Moreover, including parametric or semi-parametric learners in addition to non-parametric ones allows us to incorporate some knowledge about the data-generating process \citep{Keil_Edwards_2018}. Recent applications of SL include seasonal influenza hospitalisation \citep{Gantenberg_2023}, mortality in intensive care units \citep{Pirracchio_2015}, and prison violence \citep{Bacak_Kennedy_2019}. SL was recently expanded to joint outcome modelling \citep{Tanner_2021, Devaux_2022} and time-series \citep{Benkeser_2018, Malenica_2023}. 

Personalised predictions are motivated by the need for tailored predictions for a specific individual instead of predictions that are good on average in a population. Current personalised prediction models for an index individual are trained using similar individuals according to clustering algorithms or similarity metrics \citep{Sharafoddini_2017}. \citet{Wang_2022} proposed a sequential similarity metric for longitudinal settings but with only one prediction time. In contrast, the personalised online super learner (POSL) from \citet{Malenica_2023} leverages the longitudinal structure to personalise the prediction directly at the individual level under two mild assumptions without relying on identifying sufficiently similar individuals.

In this paper, we adapt POSL to predict a repeated continuous outcome dynamically, propose a new way to validate the performance of dynamic or personalized prediction models, and finally discuss the choices and challenges underpinning the use of the POSL. The paper is organised as follows. We introduce the motivating application in the next section. The third section describes the POSL methodology in our longitudinal context. We describe the procedures for model assessment in the fourth section and report the results from the application in the fifth section following the TRIPOD+AI statement \citep{Collins_2024}. We discuss the choices and challenges in the sixth section before concluding in the last section.

\section{Motivating application: prediction of the individual convection volume for hemodiafiltration}\label{sec2}

End-stage kidney disease is an important cause of morbidity and mortality worldwide, with a prevalence estimated at almost 5 million individuals by 2030 \citep{Liyanage_2015}. The only cure is kidney transplantation, but there is a severe shortage of transplant kidneys. In 2021, approximately one-third of transplant candidates received a kidney transplant in Canada, while the others remained on a renal replacement therapy \citep{CIHI}. 

Amongst the different renal replacement therapies, we focus on hemodiafiltration (HDF), which combines diffusive clearance and convective removal of solutes. The primary outcome of HDF is the convection volume, defined as the sum of the replacement volume and the intradialytic net weight loss achieved over the entire HDF session. Recent randomised trials and an individual patient data meta-analysis suggested that one should aim for a convection volume of at least 23 or 24L per session to reach the highest levels of efficacy and thus both survival and quality of life \citep{convince_2023, Levesque__2015, Peters_2016}. Achieving high convection volumes time after time and across patient populations requires clinical skills and readily accessible information and data. Nephrologists must continually re-assess multiple parameters refreshed with each HDF session and consider time-varying clinical status changes, which is daunting in busy dialysis centres. Therefore, a dynamic score predicting the convection volume, updated for each new HDF session, could help nephrologists focus on sub-optimal sessions and manage them by enabling a discussion with the patient about the options for improving the convection volume. For example, one may extend the session length or the needle's width \citep{Chapdelaine_2015}.
 
We illustrate the POSL through the development of a personalised dynamic prediction framework for people undergoing chronic HDF in the Centre Hospitalier de l'Université de Montréal, Québec, Canada, to predict the individual convection volume with prediction made prior to starting each HDF session.
We used an open cohort of patients undergoing chronic HDF at the outpatient dialysis clinic or the Centre externe De Gaspé. We considered HDF as chronic if there were at least 28 consecutive sessions. Patients with chronic HDF were followed from March 1\textsuperscript{st} 2017 to December  1\textsuperscript{st} 2021. Patients were permanently censored if they died, underwent renal transplantation, or were transferred to another dialysis facility. Table S1 (available online) lists the study variables with their misclassification and measurement error risks. The session-specific predictors are the following: year, season, time since the last session, haemoglobin, albumin, central venous catheter changes, intercurrent hospitalisation, dalteparin dose, access by central venous catheter, and excess weight. The baseline predictors are time since the first chronic HDF session, age, gender, hypertension, diabetes, peripheral vascular disease, congestive heart failure, arrhythmia, acute myocardial infarction, chronic pulmonary disease, liver disease, valvular disease, cancer, metastatic solid tumour, cerebrovascular disease, dementia, peptic ulcer disease, hemiplegia or paraplegia, and rheumatic disease. We also considered some individual recent sessions' history as predictors: the outcomes (convection volumes) averaged over the three last months, the alteplase dose averaged over the last week, and the number of inversion lines averaged over the last week.

\section{Personalised super learner}\label{sec3}

\subsection{Super learning}

To introduce the SL as described in \citet{vdl_Polley_Hubbard_2007}, we first consider a time-fixed setting. Let $\mathcal{O}=(\boldsymbol{X},Y)$ denote the matrix of observed variables, where $\boldsymbol{X}$ is the vector of baseline predictors, and $Y$ is the outcome. Let also $\boldsymbol{K}$ denote the set of $C$ candidate learners, indexed by $c=1,...,C$. A learner is a function taking as inputs $\boldsymbol{X}$ to provide a prediction $\hat Y$. We use uppercase letters for random variables, lowercase letters for possible values of variables or constants, bold letters for vectors, and calligraphic font for matrices. 

The SL takes three inputs: the observed dataset with $n$ draws of $\mathcal{O}$, the user-specified set of candidate learners $\mathbf{K}$, and a loss function (or a performance measure) for quantifying the prediction error. The set $\mathbf{K}$ should include different algorithms (\textit{e.g.}, tree-based, regression-based, parametric, non-parametric). It can also include multiple versions of the same algorithm with different tuning parameters or different subsets of data \citep{Phillips_2023}. 

The SL involves fitting each candidate learner to the data using cross-validation (CV) to ultimately determine either the best learner or a weighted combination of learners. The first option, the discrete SL (dSL), selects the best-performing learner amongst the $C$ candidate learners. The latter option is an ensemble SL (eSL), producing the best combination (w.r.t. the selected loss function; described below) of the predictions of the $C$ candidate learners. 

First, CV proceeds as follows. The $n$ individuals are randomly partitioned into $V$ folds. Each fold defines a specific validation set, while the remaining observations define the related training set. Each of the $C$ candidate learners is trained on each of the $V$ training sets, and $C$ predictions (one per learner) are obtained for each individual in the corresponding $V$ validation sets, resulting in the $n \times C$ matrix $\hat{\mathcal{Y}} = \biggl(\begin{smallmatrix}
\hat{\boldsymbol{y}}_{11} & \cdots & \hat{\boldsymbol{y}}_{1C} \\ 
\vdots & \ddots & \vdots \\ 
\hat{\boldsymbol{y}}_{V1} & \cdots & \hat{\boldsymbol{y}}_{VC}
\end{smallmatrix} \biggr)$, where $\hat{\boldsymbol{y}}_{vc}$ is the vector of predictions using learner $c$ for all individuals in the validation fold $v$. We thus define the meta-level dataset $\mathcal{M}=(\hat{\mathcal{Y}},Y)$, which includes the matrix of outcome predictions and the corresponding observed outcomes.   

Second, the risk with respect to the selected loss function is empirically optimised over $\mathcal{M}$. The dSL is identical to the candidate learner with the best cross-validated performance (\textit{i.e.}, yielding the lowest estimated risk (lowest average loss) in $\mathcal{M}$). The eSL uses a meta-learner to obtain a vector of length $C$ of optimal weight values (\textit{e.g.}, coefficients from a constrained linear regression) $\boldsymbol{\alpha}$, one coefficient per candidate learner, which is determined by finding the values of $\boldsymbol{\alpha}$ that optimise the expected value of the selected loss function over all individuals. 

Third, the $C$ learners are fit on the whole observed dataset $\smallO$ to obtain $C$ $n$-vectors of predictions.  

Last, we obtain the final prediction from the fitted values obtained in the previous step. The dSL prediction is that of the candidate learner selected in the second step. The eSL prediction is the linear combination of the predicted outcomes from each candidate learner using the coefficient values estimated with the meta-learner. See \citet[Figure 2]{Phillips_2023} for a visual illustration.

\subsection{Personalised online super learning for longitudinal prediction}

POSL \citep{Malenica_2023} leverages two sources of information: the individual $i$ for which we want to predict the outcome (hence \textit{personalised}) dynamically in real-time (or \textit{online}) and an “historical” pool of $H$ individuals used for the prediction, indexed by $h$ ($h=1, \ldots, H$). Suppose nephrologists want to predict the HDF convection volume for a given patient. This specific patient will be denoted by $i$. The $H$ other patients previously dialysed constitute the historical pool. The task can be viewed as an external prediction in a cross-sectional setting, where $i$ is the individual for whom the prediction is made, with a score developed using the $H$ historical individuals in addition to patient $i$'s past information. 

We denote the matrix of the $P$ time-varying predictors measured until time $t=1,...,T$ by $\mathcal{L}_i=\biggl(\begin{smallmatrix}
{L_i}_{11} & \cdots & {L_i}_{1P} \\ 
\vdots & \ddots & \vdots \\ 
{L_i}_{T1} & \cdots & {L_i}_{TP}
\end{smallmatrix} \biggr)$. Owing to the outcome being repeatedly measured over time, we denote it by $\mathbf{Y}_i=(Y_{i,1}, \ldots, Y_{i,T})$. Let $\mathcal{O}_i=(\boldsymbol{X}_i,\mathcal{L}_{i,t\leq\tau},\boldsymbol{Y}_{i,t<\tau})$ denote the observed data for the unique individual $i$ with which one wants to predict $Y_\tau$, with $t$ being the time of the observation and $\tau$ being the time of prediction. Thus, $\mathcal{L}_{i,t\leq\tau} \text{ and } \boldsymbol{Y}_{i,t<\tau}$ are the time-varying predictors matrix and the outcome vector with data up to at the last session before the predicted one, respectively.

Let $\{\mathcal{O}_h\}^H_{h=1}$, where $\mathcal{O}_h=(\boldsymbol{X}_h,\mathcal{L}_{h},\boldsymbol{Y}_{h})$, denote the observed data for the historical pool of individuals. Each individual $h$ has a series of measurements indexed by $t=1,\cdots,T_h$, a time-varying predictor matrix of the same $P$ predictors $\mathcal{L}_h=\biggl(\begin{smallmatrix}
{L_h}_{11} & \cdots & {L_h}_{1P} \\ 
\vdots & \ddots & \vdots \\ 
{L_h}_{T_h1} & \cdots & {L_h}_{T_hP}
\end{smallmatrix} \biggr)$, and a vector of historical outcomes $\boldsymbol{Y}_h=(Y_{h,1}, \ldots, Y_{h,T_h})$.  

Leveraging the longitudinal structure of the data, POSL predicts $Y_{i,\tau}$ using \textit{both} the covariates measured at previous times ($t<\tau$; $t=1, \ldots, T$) from the individual $i$ and the entire trajectory of the individuals $h=1,...,H$ (even if $t\geq \tau$ since $\tau$ does not refer to calendar time). \citet{Benkeser_2018} showed that (i) assuming the $Y_{i,\tau}$ is independent of previous observations given some fixed-dimension summaries of the past data for individual $i$ and (ii) assuming $\mathcal{O}_i$ share a common conditional distribution according to the previously defined summaries is sufficient to make the usual i.i.d. assumption.

Two sets of candidate learners can be defined depending on the set of individuals used for training. Let $\boldsymbol{K^{ind}}$ and $\boldsymbol{K^{hist}}$ be the sets of $C^{ind}$ and $C^{hist}$ candidate learners trained on the individual $i$ or individuals $h=1,...,H$, respectively. The historical learners $\boldsymbol{K^{hist}}$ use more information and provide more data for predicting the earlier outcomes. In contrast, the individual learners $\boldsymbol{K^{ind}}$ can be more accurate at later times since they are based on personal trajectory. $\boldsymbol{K^{ind}}$ and $\boldsymbol{K^{hist}}$ can include the same learners or different ones. 

Contrary to the SL described previously, POSL takes four inputs: the observed datasets $\mathcal{O}_i$, the observed historical dataset $\{\mathcal{O}_h\}^H_{h=1}$, the user-specified set $\boldsymbol{K}=(\boldsymbol{K^{ind}}, \boldsymbol{K^{hist}})$ of $C$ candidate learners ($C=C^{ind}+C^{hist}$), and a loss function for quantifying the prediction error. Note that $\{\boldsymbol{O}_h\}^H_{h=1}$ is not strictly required if $\boldsymbol{K^{hist}}$ is already trained.

First, the historical learners $\boldsymbol{K^{hist}}$ are trained on the historical cohort $\{\mathcal{O}_h\}^H_{h=1}$ if it has not yet been done. Second, the individual data $\smallO_i$ are partitioned into $V$ folds according to specific CV schemes described in the next subsection. 
The $V$ training sets are used to train the $C^{ind}$ individual learners only. Then, one obtains $C$ predictions (one per learner, whether individual or historical) for the individual $i$ in the corresponding $V$ validation sets. All of the predictions from the $V$ validation sets are merged to define the meta-level dataset $\mathcal{M}=(\hat{\mathcal{Y}}_i, \boldsymbol{Y}_{i, t<\tau})$, where $\hat{\mathcal{Y}}_i$ is the matrix of learner-and-fold-specific predictions as in the single time-point SL. Note that $\hat{\mathcal{Y}}_i$ includes predictions from both individual and historical learners such that $\hat{\mathcal{Y}}_i=(\hat{\mathcal{Y}}_i^{ind}, \hat{\mathcal{Y}}_i^{hist})$, where $\hat{\mathcal{Y}}_i^{ind}$ and $\hat{\mathcal{Y}}_i^{hist}$ denote the matrix of fold-specific predictions from $\boldsymbol{K^{ind}}$ and $\boldsymbol{K^{hist}}$, respectively. Third, $\mathcal{M}$ is used for optimising the estimated risk based on the selected loss function. The dSL is the candidate learner that minimises the risk. As in the single time-point SL, the eSL uses a meta-learner to obtain a vector of $C$ weights $\boldsymbol{\alpha}=(\boldsymbol{\alpha}^{ind}, \boldsymbol{\alpha}^{hist})$ that minimise the risk over $\mathcal{M}$. For instance, one can use as meta-learner a convex non-negative least squares (NNLS) regression formalised as follows: $\hat{\boldsymbol{Y}}_{i,t<\tau}=\boldsymbol{\alpha}\hat{\mathcal{Y}}_i$, such that $\sum\boldsymbol{\alpha}=1$, $\boldsymbol{\alpha}\geq0$, and the statistical unit is the person-time. Note that the non-negative coefficients are estimated first and then scaled with respect to the convex constraint. The related loss function is the cumulative squared error defined as $\sum_{v=1}^{V}(Y_{i,v}-\hat y_{i,v})^2$, where $Y_{i,v}$ and $\hat y_{i,v}$ are respectively the observed outcome and the prediction obtained (i.e. $\hat y_{i,v}$ is the $v$th element of $\hat{\boldsymbol{Y}}_{i,t<\tau}$) in the validation fold $v$. The meta-learner and the loss function can be weighted to give more importance to recent times over older ones. For instance, \citet{Malenica_2023} suggested setting time-specific weights $\boldsymbol{\omega}$ to 1 for the most recent time and the others to $(1-\delta)^{\tau-t}$, where $\delta$ is a user-defined constant representing how quickly past times should be down-weighted. Fourth, the candidate learners $\boldsymbol{K}$ are fit on the trajectory of the individual $i$ up to time $\tau$ to obtain $C$ predictions. Last, we obtain the final prediction from the fitted values obtained in the previous step. The dSL prediction is that of the candidate learner minimising the loss computed in step 2. The eSL prediction is the prediction from the meta-learner taking the fitted values (fourth step) as inputs. Therefore, the prediction evolves over time and will be based on the - possibly different - best learners at each time.

\subsection{Rolling cross-validation}

\begin{figure}[ht]
    \centering
    \includegraphics[width=0.8\textwidth, trim=0cm 17cm 0cm 0cm, clip=true]{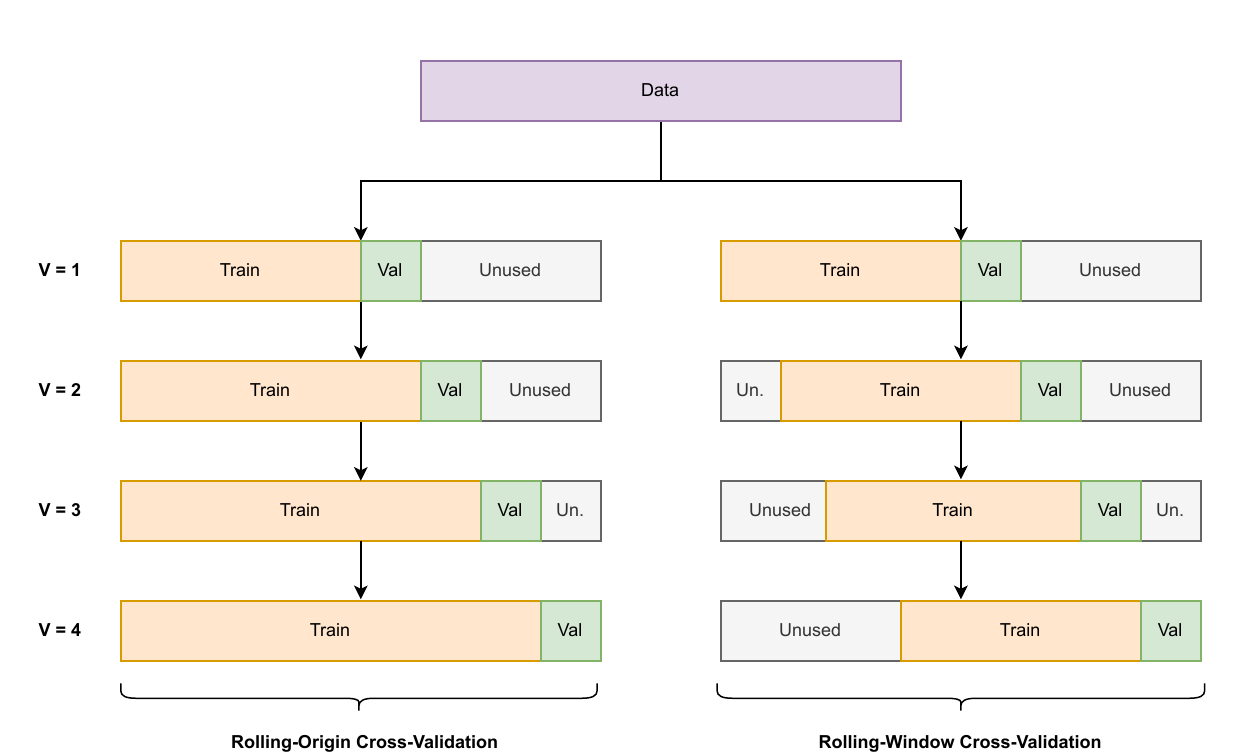}
    \caption{Rolling cross-validation schemes across four folds $v$.\\ Abbreviations: Un., Unused; and Val, Validation.}
    \label{f:cv}
\end{figure}

Owing to the temporal dependency in $\mathcal{O}_i$, the CV procedure cannot randomly assign person-time observations to either the training or validation set. One must avoid using future observations when training individual learners because these data do not yet exist to make the prediction. Therefore, instead of using a $V$-fold CV scheme, one must use a hold-out CV scheme \citep{Tashman_2000} where a subset of the timeline of data is reserved for validating the candidate learners (Figure \ref{f:cv}).   

The global idea is to start with a small subset of data as the first training fold and validate in the following time (first validation fold). The validation fold is then iteratively added to the training fold to define the next training fold, and the subsequent time becomes the new validation fold. We refer to this scheme as Rolling-Origin CV (ROCV) according to several authors \citep[\textit{e.g.},][]{Malenica_2023, Hyndman_2018}, though this term has been used to describe different CV schemes \citep[\textit{e.g.},][]{
Tashman_2000}. Alternatively, one can prune the oldest time at each update for “cleaning out” patterns from one fold to the next, obtaining a so-called rolling-window CV (RWCV). See \citet{Swanson_White_1997} for a discussion on the advantages and pitfalls of RWCV.

\section{Development and validation of the POSL for predicting HDF convection volumes}\label{sec4}

\subsection{Model development}
Since the strength of the SL is based on the diversity of the set of candidate learners \citep{Phillips_2023}, we used five algorithms with two different sets of predictors as both individual and historical learners: an unpenalised linear model with main terms only, ridge linear model, lasso linear model, multivariate adaptive regression splines (MARS), and extreme gradient boosting trees (XGBoost). Thereby, we included two different implementations of each algorithm as candidate learners: the first with all predictors and the second with only preliminary step of predictor screening with a random forest importance measure \citep{Breiman_2001b}. In addition, we considered the arithmetic mean of the outcome as a candidate learner to further reduce the risk of overfitting \citep{Balzer_Westling_2021}. We refer the readers to \citet{James_Witten_Hastie_Tibshirani_2021} for an introduction to these approaches. Software and related references are presented in Table S2. The individual learners are trained through ROCV and RWCV (with a window size of 10 sessions), doubling the number of individual learners. Note that a fold is disregarded when at least one candidate learner did not converge \citep{Malenica_2023}.

We selected a squared error loss function (or its cumulative sum for the dSL), leading to an NNLS meta-learner for the eSL \citep{Tanner_2021}. Two meta-learners were considered (yielding two different eSLs): a convex and a non-convex NNLS. Convex NNLS is recommended for SL, but \citet{Malenica_2023} reported better results with non-convex NNLS, likely because the convex combination was enforced after estimating the coefficients rather than as part of the estimation process itself. The meta-learners were weighted according to the time as described in the previous section. Specifically, the weight $\omega_t$ assigned to the squared loss measured at time $t$,  was defined as

\begin{equation*}
    \omega_t = \begin{cases}  
                    1 & \text{if } t \geq \tau-5 \\
                    (1-0.1)^{\tau-t} & \text{otherwise.}    
                \end{cases}   
\end{equation*}

We truncated the predictions within $[0,50]$ to respect clinically plausible values.

\subsection{Model validation}
\begin{figure}[ht]
    \centering
    \includegraphics[width=0.8\textwidth]{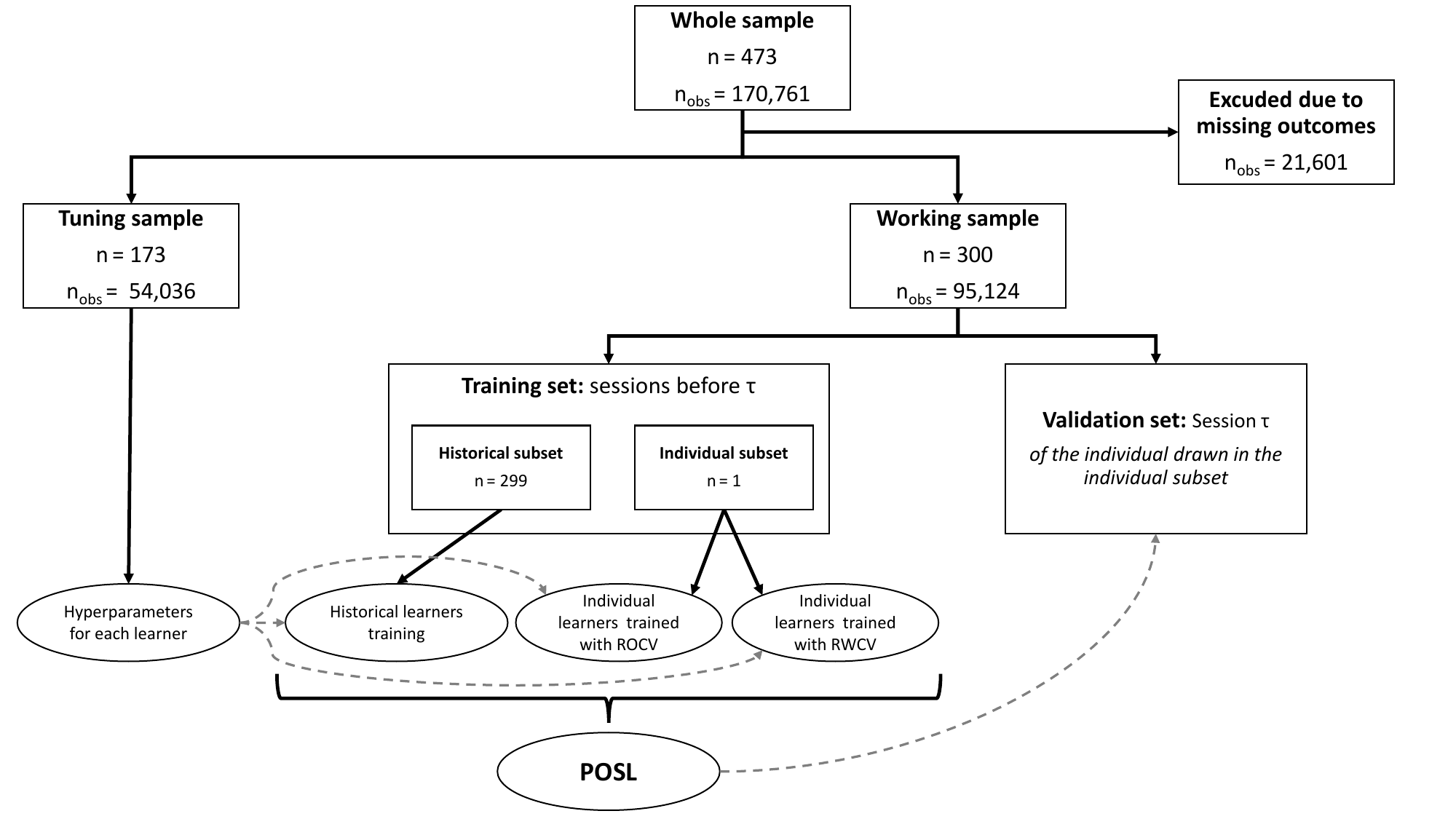}
    \caption{Flow-chart of the tuning-training-validation process.\\ Abbreviations: $n$, Number of individuals; $n_{obs}$, Number of sessions; ROCV, Rolling-Origin Cross-Validation; RWCV, Rolling-Window Cross-Validation; POSL, Personalised Online Super Learner; and $\tau$, session for training POSL.}
    \label{f:set}
\end{figure}

We split the whole sample into tuning and working samples. The former was used to define, through a 10-fold CV, the learners’ hyperparameters used in the latter. We applied the model development procedure in the working sample for each individual (but one at a time). Therefore, $\smallO_i$ corresponds to one individual and $\{\smallO_h\}$ to the remaining individuals. We summarise the sample-splitting process in Figure \ref{f:set}). 

\begin{figure}[ht]
    \centering
    \includegraphics[width=0.8\textwidth]{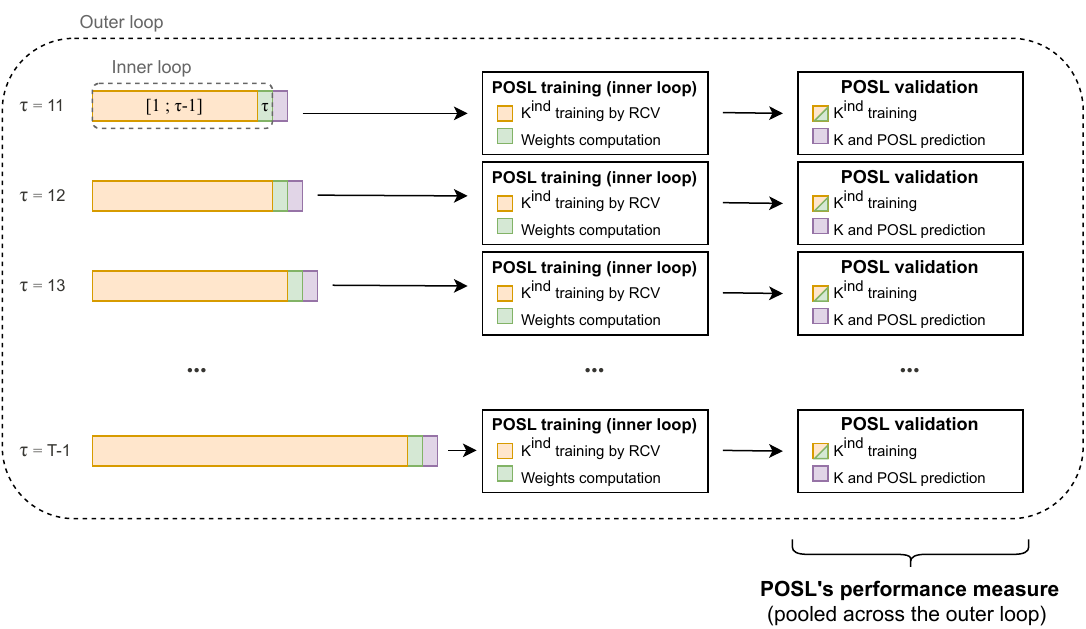}
    \caption{POSL's forward cross-validation.\\ Abbreviations: $K$, Candidate learners set; $K^{ind}$, Candidate individual learners set; RCV, Rolling cross-validation (either origin or window); POSL, Personalised Online Super Learner; $T$, last session; and $\tau$, session for training POSL. Adapted from \citet{Li_2021}}
    \label{f:forw}
\end{figure}

To validate the prediction made by the POSL, we used a nested validation scheme called forward validation \citep{Hjorth_1982}. Assuming we are interested in predicting the session's outcome $\tau+1$, the sessions from 1 to $\tau$ are used as a training set for training the POSL with its own loop of ROCV or RWCV. Then, the session $\tau+1$ is the validation set in which we predicted the outcome (Figure \ref{f:forw}). Predictions were made every session from the 12\textsuperscript{th} ($\tau=11$ gives ten sessions for training the POSL). Importantly, no session appears simultaneously in the training and validation sets, limiting over-fitting.

\subsection{Model performance}
For each individual $i$, we computed two accuracy measures \citep{Hyndman_2018}: the median absolute error (MdAE) and the mean squared error (MSE) defined, respectively, as $MdAE_i=\text{median}_{t=1}^{T_i}(\lvert y_{ti}-\hat y_{ti} \rvert)$ and $MSE_i=T_i^{-1}\sum_{t=1}^{T_i}(y_{ti}-\hat y_{ti})^2$, where $\hat y_{ti}$ and $y_{ti}$ are the prediction and the actual outcome at time $t$ for the individual $i$, respectively. Note that the MSE is sometimes called Brier Score in predictive modelling. We assessed individual-specific calibration through the calibration intercept, the calibration slope and a flexible calibration curve \citep{VanCalster_2019}. We used a generalised additive model with integrated smoothness estimation for obtaining the flexible curve \citep{Wood_2016}. We assessed individual-specific discrimination by first dichotomising the predictions according to a clinically relevant threshold according to the literature (specifically, $\geq$24L vs $<$24L). Then, we computed the area under the ROC curve (AUROC). Furthermore, we compared the performances of POSL to those of the candidate learners. 

In addition, we evaluated the evolution over time of the MdAE, calibration slope, and calibration intercept averaged (using the median to circumvent extreme predictions) over all individuals for POSL, the individual linear model, and the historical XGBoost. Finally, we evaluated the clinical utility by computing the net benefit, defined by $\text{net benefit} = \text{true positives rate} - \text{false positives rate} \times p/(1-p)$, where $p$ is the binarised outcome prevalence \citep{Peirce_1884}, at a range of binarisation thresholds.

\subsection{Other considerations}
Because there is no current baseline performance data, the published literature is not informative, and no minimal sample size formula was available using an ML approach, we did not calculate a minimal sample size. For this study, we used convenience samples of all patients treated with chronic HDF.

Missing predictors were imputed separately
for each individual with the median or the mode of the individual’s previous sessions for continuous and discrete predictors, respectively, avoiding temporal data leakage \citep{Roberts_2017}. Missingness indicators were added in the predictors’ set as the pattern of predictor missingness might be informative for predicting the outcome \citep{Sperrin_2020}. We excluded sessions with a missing outcome (\textit{i.e.}, either cancelled or incomplete). Table S3 presents the characteristics of the included and excluded sessions. 

We performed a sensitivity analysis with a binary outcome (at least 24L was successful; otherwise unsuccessful) where missing outcomes were imputed as unsuccessful. The procedure was almost identical, except that we evaluated the clinical utility through a decision curve analysis \citep{Vickers_Elkin_2006}, and the loss function for the dSL was the negative log-likelihood to favour calibration and accuracy against discrimination. Furthermore, the negative log-likelihood loss function may even achieve better AUROC than an AUROC-maximising loss function, especially in smaller sample sizes \citep{LeDell2016}. 

\begin{table}[h!]
\centering
\begin{minipage}{\textwidth}
\caption{Description of the tuning and working samples.\label{t:desc}}%
\begin{tabular}{lrrrcc}
  \toprule
 & Overall & Working & Tuning & \multirow{2}{*}{SMD} & \multirow{2}{*}{Missing (\%)} \\ 
 & N=149,160 & N=95,124 & N=54,036 &  \\ 
  \midrule
  Time between HDF sessions, days (mean (sd)) &    2.45 (5.39) &    2.46 (5.39) &    2.45 (5.39) &  0.003 & 0.0 \\ 
  Time in the RTWL, days (mean (sd)) &  583.07 (399.26) &  581.25 (398.40) &  586.27 (400.75) &  0.013 & 0.0 \\ 
  Hemoglobin, g/L (mean (sd)) &  108.21 (13.35) &  107.92 (12.93) &  108.71 (14.05) &  0.058 & 0.0 \\ 
  Albumin, g/L (mean (sd)) &   37.20 (3.86) &   37.17 (3.88) &   37.25 (3.82) &  0.021 & 0.6 \\ 
  Dalteparin dose, IU (mean (sd)) & 5,067.45 (2,484.22) & 4,946.77 (2,429.68) & 5,279.89 (2,563.69) &  0.133 & 0.0 \\ 
  Age at baseline, years (mean (sd)) &   67.17 (13.92) &   67.29 (13.86) &   66.96 (14.03) &  0.024 & 0.0 \\ 
  Excess weight, kg (mean (sd)) &    1.74 (1.29) &    1.72 (1.28) &    1.79 (1.32) &  0.060 & 0.8 \\ 
  Convection volume history\dag, L (mean (sd)) &   27.36 (2.91) &   27.40 (2.84) &   27.28 (3.03) &  0.040 & 0.0 \\ 
  Inversion of HDF lines history\dag (mean (sd)) &    0.02 (0.09) &    0.02 (0.09) &    0.02 (0.09) &  0.003 & 0.0 \\ 
  Alteplase doses history\dag (mean (sd)) &    0.05 (0.26) &    0.05 (0.25) &    0.06 (0.27) &  0.038 & 0.0 \\ 
  Year (N, (\%)) &  &  &  &  0.057 & 0.0 \\ 
  \textcolor{white}{ccc}2017 & 25,827 (17.3)  & 16,334 (17.2)  &  9,493 (17.6)  &  &  \\ 
  \textcolor{white}{ccc}2018 & 39,870 (26.7)  & 24,657 (25.9)  & 15,213 (28.2)  &  &  \\ 
  \textcolor{white}{ccc}2019 & 41,597 (27.9)  & 26,941 (28.3)  & 14,656 (27.1)  &  &  \\ 
  \textcolor{white}{ccc}2020 & 36,196 (24.3)  & 23,497 (24.7)  & 12,699 (23.5)  &  &  \\ 
  \textcolor{white}{ccc}2021 &  5,670 ( 3.8)  &  3,695 ( 3.9)  &  1,975 ( 3.7)  &  &  \\ 
  CVC change since last session (N, (\%)) &   388 ( 0.3)  &   250 ( 0.3)  &   138 ( 0.3)  &  0.001 & 0.0 \\ 
  Hospitalisation (N, (\%)) &  8,013 ( 5.4)  &  5,662 ( 6.0)  &  2,351 ( 4.4)  &  0.072 & 0.0 \\ 
  Access by CVC (N, (\%)) & 70,851 (47.5)  & 44,351 (46.6)  & 26,500 (49.0)  &  0.048 & 0.0 \\ 
  Male (N, (\%)) & 94,723 (63.5)  & 60,909 (64.0)  & 33,814 (62.6)  &  0.030 & 0.0 \\ 
  Hypertension (N, (\%)) & 90,435 (60.6)  & 57,891 (60.9)  & 32,544 (60.2)  &  0.013 & 0.0 \\ 
  Diabetes (N, (\%)) & 59,296 (39.8)  & 35,304 (37.1)  & 23,992 (44.4)  &  0.149 & 0.0 \\ 
  Peripheral vascular disease (N, (\%)) & 32,449 (21.8)  & 20,597 (21.7)  & 11,852 (21.9)  &  0.007 & 0.0 \\ 
  Congestive heart failure (N, (\%)) & 28,583 (19.2)  & 19,963 (21.0)  &  8,620 (16.0)  &  0.130 & 0.0 \\ 
  Cardiac arrhythmia (N, (\%)) & 28,962 (19.4)  & 18,753 (19.7)  & 10,209 (18.9)  &  0.021 & 0.0 \\ 
  Acute myocardial infarction (N, (\%)) & 22,214 (14.9)  & 15,766 (16.6)  &  6,448 (11.9)  &  0.133 & 0.0 \\ 
  Chronic pulmonary disease (N, (\%)) & 20,563 (13.8)  & 14,126 (14.9)  &  6,437 (11.9)  &  0.086 & 0.0 \\ 
  Liver disease (N, (\%)) & 17,501 (11.7)  & 11,783 (12.4)  &  5,718 (10.6)  &  0.057 & 0.0 \\ 
  Valvular disease (N, (\%)) & 15,130 (10.1)  & 10,657 (11.2)  &  4,473 ( 8.3)  &  0.099 & 0.0 \\ 
  Cancer (N, (\%)) & 12,369 ( 8.3)  &  6,788 ( 7.1)  &  5,581 (10.3)  &  0.113 & 0.0 \\
  Metastatic solid tumour (N, (\%)) &  1,943 ( 1.3)  &  1,054 ( 1.1)  &   889 ( 1.6)  &  0.046 & 0.0 \\ 
  Cerebrovascular disease (N, (\%)) & 11,904 ( 8.0)  &  7,361 ( 7.7)  &  4,543 ( 8.4)  &  0.025 & 0.0 \\ 
  Dementia (N, (\%)) &  4,931 ( 3.3)  &  3,485 ( 3.7)  &  1,446 ( 2.7)  &  0.056 & 0.0 \\ 
  Peptic ulcer disease (N, (\%)) &  4,383 ( 2.9)  &  2,041 ( 2.1)  &  2,342 ( 4.3)  &  0.124 & 0.0 \\ 
  Hemi- or paraplegia (N, (\%)) &  4,470 ( 3.0)  &  1,830 ( 1.9)  &  2,640 ( 4.9)  &  0.164 & 0.0 \\ 
  Rheumatic disease (N, (\%)) &  3,268 ( 2.2)  &  2,744 ( 2.9)  &   524 ( 1.0)  &  0.140 & 0.0 \\ 
  Week-end HDF session (N, (\%)) & 24,959 (16.7)  & 16,009 (16.8)  &  8,950 (16.6)  &  0.007 & 0.0 \\ 
  Season (N, (\%)) &  &  &  &  0.014 & 0.0 \\ 
  \textcolor{white}{ccc}fall & 37,047 (24.8)  & 23,481 (24.7)  & 13,566 (25.1)  &  &  \\ 
  \textcolor{white}{ccc}spring & 38,388 (25.7)  & 24,501 (25.8)  & 13,887 (25.7)  &  &  \\ 
  \textcolor{white}{ccc}summer & 36,513 (24.5)  & 23,468 (24.7)  & 13,045 (24.1)  &  &  \\ 
  \textcolor{white}{ccc}winter & 37,212 (24.9)  & 23,674 (24.9)  & 13,538 (25.1)  &  &  \\ 
   \botrule
\end{tabular}
\begin{tablenotes}%
\item[\dag] Averaged over three months (convection volume) or one week (inversion lines and alteplase)
\item Abbreviations: CVC, Central venous catheter; HDF, Hemodiafiltration; IU, International unit; N, Number of sessions; RTWL, Renal transplant waiting list; SMD, Standardised mean difference. 
\end{tablenotes}
\end{minipage}
\end{table}

\section{Results}\label{sec5}

\subsection{Sample description}

The study sample comprised 149,160 observations (\textit{i.e.}, sessions) for 473 individuals with, on average, 315 sessions per individual (min: 1, max: 776). The convection volume ranged from 8L to 40L, with a mean of 27.4L. Seventy-two percent of the sessions presented a convection volume greater or equal to 24L. Characteristics of the tuning and working samples are described in Table \ref{t:desc}. Note that the sample presented only a few missing data, with 0.8\% of subjects missing excess weight and 0.6\% missing albumin. Recall that three POSL implementations were considered: (a) a convex NNLS eSL, (b) a non-convex NNLS eSL, and (c) a dSL.  The POSLs did not converge for two individuals in the working sample (0.67\%) due to an insufficient number of recorded sessions (\textit{i.e.}, less than twelve). Among the 2,295,825 predictions, 1583 (0.07\%) were trimmed due to values outside of the bounds. No historical learners led to such predictions. Individual MARS (regardless of random forest screening) predicted extreme values the most often (0.50\%), followed by the individual linear model (0.22\%) and the individual lasso (0.15\%). RF screening reduced their percentage of extreme predictions to 0.054\% and 0.049, respectively. Among the POSLs, the dSL estimated less often extreme predictions than the two eSLs (0.04\% vs 0.06\%). 

\begin{figure}[h!]
    \centering
    \includegraphics[width=\textwidth, trim=0 0 0 15]{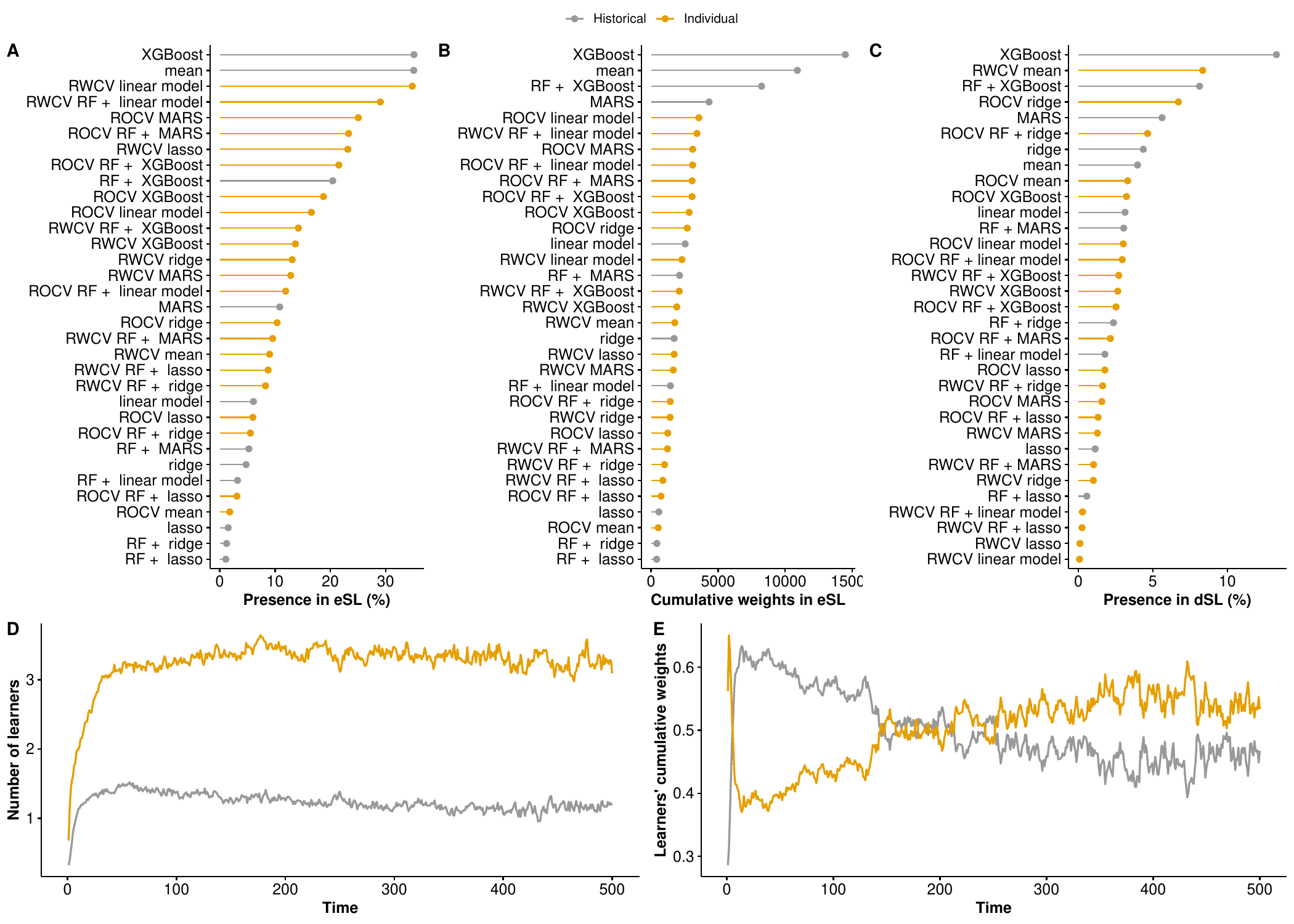}
    \caption{Candidate learner distributions in the non-convex eSL and the dSL. (A) Percentage non-null weights in the eSL, (B) Sum of weights across time and individuals, (C) Percentage selection by dSL, (D) Number of learners composing the non-convex eSL over time, and (E) Sum of weights over time averaged between individuals. The X-axis of panels D and E is bound to 500 due to an insufficient number of observations after, and time 1 corresponds to the first predicted session.\\ Abbreviations: dSL, discrete super learner; eSL, ensemble super learner; MARS, multivariate adaptive splines regression; RF, random forest screening before training; ROCV, rolling-origin cross-validation; RWCV, rolling-window cross-validation; and XGBoost, extreme gradient boosting tree.}
    \label{f:poids}
\end{figure}

\subsection{Candidate learner selection}

A nearly identical distribution of candidate learners was found between the convex and non-convex eSLs. Globally, the historical learners -- especially XGBoost -- were more important in the eSLs and the dSL than individual learners (Figure \ref{f:poids}, panels A-C). Individual learners trained through ROCV were more present in the meta-learners than those trained through RWCV. The random forest screening step did not seem to improve performance since most of these learners were rarely present in the meta-learners, a notable exception being the historical XGBoost. The historical and individual lasso were seldom selected, consistent with the lack of improved performance observed in previous simulation studies \citep{VanCalster_2020, Riley_2021}. Furthermore, the eSLs included, on average, three individual and one historical learners with a constant trend over time. However, their relative importance fluctuated over time. Historical learners had more influence in the beginning until, approximately after the 200\textsuperscript{th} session, the individual learners dominated (Figure \ref{f:poids}, panels D-E).

\subsection{Performance of POSL against its candidate learners}

\begin{figure}[h!]
    \centering
    \includegraphics[width=\textwidth]{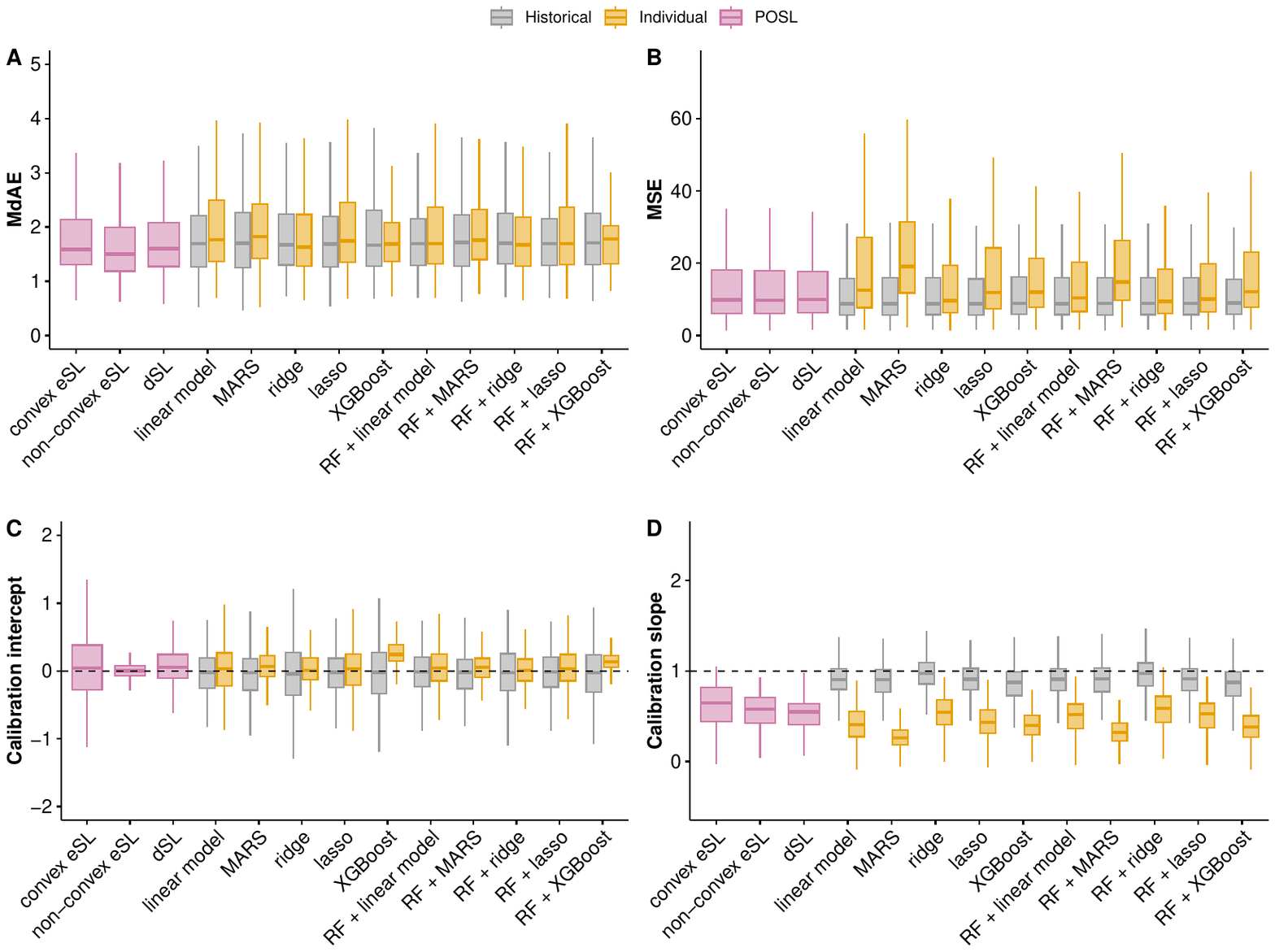}
    \caption{Time-pooled performance measures of the different POSL implementations and their candidate learners. (A) Accuracy assessed by the MdAE, (B) Accuracy assessed by the MSE, (C) Calibration-in-the-large (or mean calibration) assessed by the calibration intercept, and (D) Weak calibration assessed by the calibration slope. The dashed line represents the median of the best-performing method for accuracy measures or the ideal value for calibration measures.\\ Abbreviations: dSL, discrete super learner; eSL, ensemble super learner; MARS, multivariate adaptive regression splines; MdAE, median absolute error; MSE, mean square error; POSL, Personalised Online Super Learner; RF, random forest screening before training; and XGBoost, extreme gradient boosting tree.}
    \label{f:res}
\end{figure}

The accuracy of these three POSLs was similar (Figure \ref{f:res}), with a slight advantage for the non-convex eSL in line with previous results of \citet{Malenica_2023}. The calibration-in-the-large was near-perfect for all POSLs (\textit{i.e.}, the predictions are close to the true values on average), but the convex eSL presented a higher variability. In contrast, the weak calibration was suboptimal for all POSLs (\textit{i.e.}, there is systematic overestimation or underestimation), albeit convex eSL seemed slightly better. Historical learners almost always outperformed individual learners, especially in terms of weak calibration. Individual and historical learners were well-calibrated-in-the-large, but individual learners had a lower variance (except XGBoost). Individual MARS achieved poorer accuracy, while ridge regression yielded the best performance amongst individual learners. Additional numerical results are presented in Table S4. All historical learners were calibrated-in-the-small (\textit{i.e.}, the predictions match the true values for almost all patient strata), while individual learners and POSLs underestimated the high convection volume and overestimated the low convection volume (Figure S1 available online). Individual XGBoost yielded poorer calibration-in-the-small, especially for the lowest outcomes. The eSLs achieved the highest discrimination with an AUROC of 0.825 (95\%CI from 0.821 to 0.829) and 0.821 (95\%CI from 0.817 to 0.825) for the non-convex and the convex eSLs, respectively (Figure S2). Individual learners achieved greater AUROCs than historical learners and dSL, except for MARS and XGBoost. Random forest screening did not seem to improve any performance measures significantly. Finally, non-convex eSL achieved the best net benefit around the threshold previously identified in the literature \citep{convince_2023, Levesque__2015}, while the historical XGBoost achieved a lower net benefit than POSLs and the individual linear model regardless of the threshold (Figure S3).

\begin{figure}[h!]
    \centering
    \includegraphics[width=\textwidth]{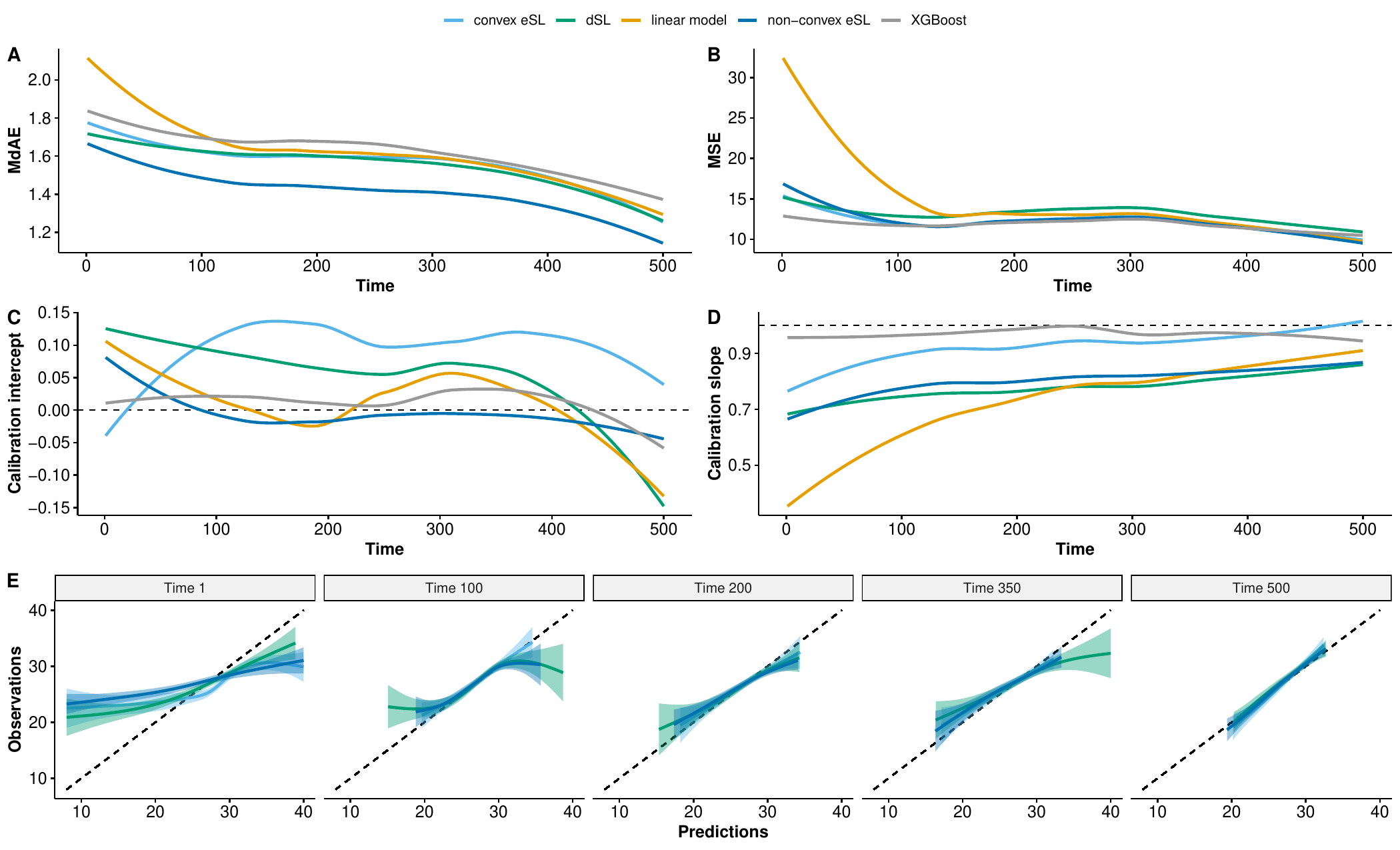}
    \caption{Performance measures of the different POSL implementations, the individual linear model, and the historical XGBoost. (A) Accuracy assessed by the MdAE, (B) Accuracy assessed by the MSE, (C) Calibration-in-the-large (or mean calibration) assessed by the calibration intercept, and (D) Weak calibration assessed by the calibration slope. (E) Calibration-in-the-small (or moderate calibration) of the POSLs at five time points. The dashed line represents an ideal calibration. X-axes of panels 1-D are bound to 500 due to insufficient observations after. Curves of panels A-D were loess-smoothed. Time 1 corresponds to the first predicted session. Flexible curves of panel E were obtained through generalised additive models with integrated smoothness estimation.\\ Abbreviations: dSL, discrete super learner; eSL, ensemble super learner; MdAE, median absolute error; MSE, mean square error; POSL, Personalised Online Super Learner; and XGBoost, extreme gradient boosting tree.}
    \label{f:tres}
\end{figure}

When looking at the evolution of the performances over time (Figure \ref{f:tres}), the individual linear model achieved poorer accuracy in the early time periods (approximately until the 150\textsuperscript{th} session). Interestingly, this time corresponds to the beginning of the equilibrium period between historical and individual learners in the eSL (Figure \ref{f:poids}E). In contrast, POSLs consistently outperformed the candidate learners regarding MdAE, the non-convex eSL achieving the best MdAE over all time points. Historical XGBoost presented the lowest MSE until the hundredth session, from which it became equivalent to the two eSLs. Similarly, the historical XGBoost achieved the best calibration-in-the-large at the beginning before being slightly overtaken by the non-convex eSL. Surprisingly the convex eSL presented the reverse pattern. For the weak calibration, historical XGBoost outperformed the other approaches until the last times, followed by the convex eSL. The individual linear model caught up with the dSL and the non-convex eSL from approximately the 250\textsuperscript{th} session, which corresponds to the time when the individual learners overtook the historical ones in the eSLs (Figure \ref{f:poids}E). POSLs were highly miscalibrated-in-the-small at the first session, but the performance improved over time with an acceptable calibration-in-the-small achieved from the 100\textsuperscript{th} session for the eSLs (Figure \ref{f:tres}E). 

\subsection{Sensitivity analysis in a classification setting}

First, note that the lasso was absent from the set of candidate learners since the best cross-validated penalisation in the tuning set was the absence of penalisation. As with a continuous outcome, the historical learners (especially XGBoost) were more present in the convex eSL than the individual learners (Figure S4). However, the relative importance of historical and individual learners was relatively constant over time, with a slight advantage for the historical learners until the 200\textsuperscript{th} session and a clearer advantage for individual learners after the 250\textsuperscript{th} session, approximately. Nevertheless, these advantages were less pronounced than those observed with a continuous outcome. In contrast, the non-convex eSL produced predictions greater than one in 13.1\% of the cases because of overly extreme weights for the individual logistic models. All of these predictions were bound to one for computing the performance measures.  

Regarding performance (Figures S5-S6, Table S5), historical learners achieved poorer MdAE and weak calibration than individual learners, but a better MSE. POSLs outperformed the historical learners regarding the MdAE and weak calibration but only outperformed the individual learners regarding the MSE. All methods achieved a correct and similar calibration-in-the-large. Between the POSLs, the non-convex eSL seemed slightly better, especially regarding the MdAE. In contrast, historical learners yielded the best calibration-in-the-small, while individual learners underestimated the convection volume more largely than the POSLs. Historical learners achieved a better AUROC than individual learners, solely outperformed by the eSLs (Figure S7). All POSLs presented a similar improved net benefit against the strategy of closely monitoring all patients when they are at least 25\% chance of having a successful HDF session (Figure S8). The evolution of the performance over time of POSLs is relatively constant (Figure S9), although the eSLs were miscalibrated-in-the-small in the first time.

\section{Choices and challenges}\label{sec6}

\subsection{To combine or not to combine?}

This paper describes the implementation of POSL \citep{Malenica_2023} to dynamically predict the individual HDF convection volume using individual trajectories and data available from a historical pool of other individuals. Furthermore, POSL can combine multiple parametric and non-parametric algorithms to obtain the best combination of their predictions (or the best algorithm's predictions) at each time. We have studied three POSLs: a dSL (the best candidate learner in terms of cumulative squared error), a convex NNLS eSL, and a non-convex NNLS eSL. We showed that their performances were close in our case, although the non-convex eSL was more discriminant and slightly more accurate than the two others. In addition, the non-convex eSL achieved correct calibration faster than the dSL in our context. These results confirm the previous findings of \citet{Malenica_2023} about the weakness of imposing a convex scaling after estimating the NNLS coefficients.

Although asymptotic theory demonstrates that the SL outperforms its candidate learners in both i.i.d. \citep{Dudoit_vdL_2005, vanderVaart_2006} and non-i.i.d. \citep{Benkeser_2018} settings, some candidate historical learners outperformed the POSLs regarding MSE and weak calibration in this application in our finite sample application. Nevertheless, \citet{Hibon_Evgeniou_2005} argued, “\textit{the advantage of combining [predictions] is not that the best possible combinations perform better than the best possible individual [prediction], but that it is less risky in practice to combine [predictions] than to select an individual [prediction] method.}”. Indeed, investigators don't know before the analysis which algorithm performs the best; using POSL ensures that they do not pick an inefficient algorithm, albeit POSL may not perform the best in all cases. We consider this as the main strength of using POSL over its alternatives (\textit{i.e.}, using a sole candidate learner or algorithms averaging approaches without a validation step \citep[Table 1]{Dormann_2018}).

\subsection{Which inputs to use for the POSL?}

Like any ensemble method, POSL's performance depends on its three inputs: (a) the data, (b) the library of candidate learners, and (c) the loss function.

For the first input, the dynamic nature of POSL requires careful thinking about the choice of predictors. In particular, lagged values of crucial predictors must be defined according to domain knowledge. We used summaries of the recent history of the outcomes and two session-specific predictors according to clinical knowledge. Nevertheless, the history of graft rejection was unavailable from our data source, which might constitute a limitation of this study. One must also keep in mind that the prediction model is developed for being \textit{implemented and used} in practice. For instance, one might avoid including predictors that are difficult to collect in an automatic way \citep{Vickers_Cronin_2010}. Similarly, predictors collected differently over time may induce model miscalibration \citep{Luijken_2020} even if RWCV might partially correct this. We used a window of 10 sessions to be able to predict sooner. However, increasing the window size may improve the RWCV-based candidate individual learners' performance by limiting overfitting. We previously noted that no minimal sample size could be derived for such algorithms. Therefore, one must assume that the collected sample is sufficient for training the algorithms, which is a strong assumption since ML algorithms are data-hungry \citep{vanderPloeg_2014}. Furthermore, an \textit{efficient} sample size must also be computed for defining a minimum number of CV folds \citep{Phillips_2023}. For the POSL, it is unclear whether these sample sizes refer to the number of individuals or the number of observations for the individual $i$, and whether we should include the historical cohort in the calculation. In our application, we have observed that the individual learners became relevant from, on average, the 150\textsuperscript{th} observation approximately and required around a hundred additional observations to be predominant. Another issue is the presence of missing data. We used a missingness indicator approach with median imputation because it is computationally non-intensive, and missing data patterns may be predictive by themselves \citep{Sperrin_2020}. However, knowing that the missingness of a predictor is predictive may motivate the user to refrain from measuring that predictor, creating harmful feedback between the model and the user \citep{Groenwold_2020}. Since POSL is a ”black box”, this phenomenon may be less likely to occur. However, one may sometimes want to know the contribution of all predictors, mainly for fairness reasons \citep{Basu_2023}. \citet{Taghizadeh_2021} described a permutation feature importance analysis \citep{Fisher_2019} to rank the predictors in an SL. Although generally considered the gold standard, multiple imputation is difficult to use with POSL because it substantially increases the computational complexity. An under-exploited alternative could be relying on ML algorithms that naturally handle missing data, such as RF or XGBoost \citep{Nijman_2022}. However, including only tree-based algorithms in the POSL restricts its flexibility and increases the risk of overfitting \cite{Balzer_Westling_2021}. Finally, the missing predictors' imputation can be viewed as a preliminary step in the SL. Thereby, different imputation schemes could be tested in the SL, such as the candidate learners with missing indicators and the same candidate learners with a single imputation.

Considering the choice of candidate learners, following the recommendation of \citet{Phillips_2023}, we incorporated a diverse library of candidate learners, including parametric models with and without penalisation, non-parametric algorithms that do not enforce linear or monotonic outcome-covariate relationships, and the arithmetic mean. Moreover, we considered two different CV schemes for individual learners to limit the impact of potential distribution shifts. Finally, we added an RF screening procedure to reduce the number of predictors to five. It is possible to pick a larger number of predictors, but there is a tradeoff between the window size and the number of time-varying predictors. With a small window size for the individual learner's cross-validation, the time-varying predictors may sometimes remain steady due to the small number of observations. However, \citet{Strobl_2007} showed that RF screening might incorrectly favour the selection of continuous/multinomial predictors over binary ones in classification settings. In the context of POSL, it could be relevant to add time-series models, such as ARIMA \citep{Hyndman_2018}, to the pool of individual learners when the number of observations is sufficient. However, including too many candidate learners (as evoked above) might complicate the estimation of the meta-learner by falling into the curse of dimensionality. Indeed, the meta-level dataset could present many more predictors (\textit{i.e.}, the candidate learner-specific predicted outcomes) than observations. Using a meta-screener before fitting the meta-learner could be a solution for reducing the dimensionality (\textit{e.g.}, using a post-lasso NNLS \citep{Belloni_Chernozhukov_2013} as meta-learner). Alternatively, we can put different eSLs and their candidate learners in a dSL \citep{Phillips_2023}, which will select a candidate learner over the eSL if the eSL does not converge. Nevertheless, this procedure requires an additional loop of CV to compute the eSL-loss needed by the dSL. \citet{Phillips_2023} defined an efficient sample size of 500 as a rule of thumb for including as many learners as computationally possible in an SL. Further research is needed to assess this rule with POSL. SLs can also be considered as candidate learners, a historical SL being especially appealing for improving the first POSL predictions. However, this requires including a supplementary loop of CV in the POSL training process. The choice of the candidate learners implies defining their hyperparameters. Indeed, default values provided by statistical packages are unlikely the optimal ones \citep{Probst_2019}. We applied a sample-splitting strategy by defining the hyperparameters and training the algorithms on mutually exclusive splits. We fixed our hyperparameters across time, which is unlikely to be optimal. Some values of the individual learners' hyperparameters may have been set too aggressively regarding the amount of data available in the first CV folds, explaining the MARS' overfitting trend. Alternatively, one can compute time-specific hyperparameters by splitting the CV fold in tuning and training mutually exclusive sets \cite{Tashman_2000}. Such a nested CV has been shown to improve performance by \citet{Varma_Simon_2006}, while other authors advocated for the more pragmatic approach of estimating the hyperparameters only one time to reduce the computational burden at the cost of a small decrease in performance \citep{Schumacher_2007, Foucher_Danger_2012}. Another solution might be to include several versions of the same candidate learner but with different hyperparameter values. Thereby, POSL can select the best implementation at each time.

The final component is the loss function. We used the sum of the squared errors over time for the dSL and the related NNLS as a meta-learner for the eSLs. \citet{Krikella_2024} suggested a decomposition of the Brier score as a loss function, weighted to emphasising either the discrimination or the calibration-in-the-large. Note that non-parametric algorithms can be used as alternative meta-learners \citep{Phillips_2023}. Our loss function and meta-regressions were weighted to give less importance to the least recent sessions. We used weights similar to those proposed by \citet{Malenica_2023}, but they acknowledged that domain knowledge must drive this choice. Again, we could define several eSLs with different weighting schemes to put into an unweighted dSL. In addition, the predictions from the CV folds were not bounded to plausible values, which potentially biased the squared error at time $t$ but also at the subsequent times since the loss function was the cumulative sum. This can explain why the dSL did not achieve the best MSE in our application. Using this loss function also implied disregarding the fold in which at least one learner did not converge to not bias the cumulative loss.

\subsection{How to evaluate the POSL}

Prediction models must be evaluated regarding overall performances, calibration, discrimination, and clinical usefulness \citep{McLernon_2023}. Calibration refers to how well model predictions agree with the actual outcome frequencies in the population under study. Of note, calibration was seldom reported for ML-based prediction models \citep{Andaur_2022}. Discrimination refers to how well the model predictions separate high- from low-risk patients. Overall performances mix calibration and discrimination. Clinical usefulness refers to how well the model predictions can improve clinical decision-making by targeting high-risk patients for additional intervention while not burdening the system and patients with excessive false positives. We used both the MSE and the MdAE as measures of overall performance. Nevertheless, these measures are time-scale dependent. If the time range is highly variable between individuals, one can prefer scale-independent measures such as the median absolute percentage error or the mean squared percentage error \citep{Hyndman_2018}. Other metrics can be considered, especially proper scores like the Brier score (albeit equivalent to the MSE with a binary outcome) or the continuous ranking probability score when the interest lies in a cumulative distribution function \citep{Tyralis_Papacharalampous_2024}. Calibration was assessed at three nested levels \citep{VanCalster_2019}. First, calibration-in-the-large implies that the prediction is close to the observation. Second, weak calibration implies no overestimation or underestimation of the outcome. Third, calibration-in-the-small implies the predictions correspond to the observations (but not necessary for each predictor pattern). We used the calibration intercept, the calibration slope, and a flexible calibration curve, respectively. Discrimination and clinical usefulness are more difficult to evaluate with a continuous outcome. Indeed, clinical decision-making typically operates in a classification setting with a threshold defining high-risk and low-risk subgroups. We used a threshold of 24L according to the literature \citep{Levesque__2015, convince_2023} for the discrimination evaluated through the AUROC. Clinical usefulness can be evaluated through the net benefit solely or a decision analysis curve for a binary outcome \citep{Vickers_Elkin_2006}. Although decision curve analysis allows for investigating the net benefit across a range of possible threshold probabilities, it should not be used to select the best threshold \citep{Vickers_2019}. Note that if the envisaged intervention is dangerous or costly, we can incorporate this “harm” in the net benefit computation \citep{Vickers_Elkin_2006}. We emphasise the importance of reporting all these performance measures to compare the algorithms fairly. Indeed, historical learners were better calibrated than individual learners and POSLs in our application, but they achieved lower discrimination and net benefit. These performance measures should be considered together, and their respective importance should be considered with respect to the intended use \citep{Vickers_Cronin_2010}. Another way to evaluate prediction models' performance, mainly suggested in the ML literature, is by providing prediction intervals instead of point prediction. Therefore, one can derive discrimination-like measures by checking if the actual outcomes lie in the intervals. However, weak calibration and calibration-on-the-small are more challenging to assess. Two main frameworks may be used to obtain prediction intervals in dynamic settings: moving block bootstrap \citep{Ju_2015} and conformal prediction \citep{Zaffran_2022}. Nevertheless, these procedures are data-hungry and time-consuming, especially with POSL.  

The traditional paradigm in prediction modelling involves developing a prediction model in a dataset, validating its performances internally on the same dataset with a correction for overfitting (\textit{e.g.}, through V-fold CV or bootstrap resampling), and validating externally in another population \citep{Steyerberg_Harrell_2016}. The cost of personalising the prediction with POSL, because POSL is based on individual learners (\textit{i.e.}, learners trained only on the previous data of the individual for which one wants a prediction), is that no general prediction model is developed, making an external validation infeasible. Nevertheless, note that prediction models are seldom appropriately validated because this needs many external validations in various populations, including those of the predicted individual \citep{VanCalster_2023}. We can only validate it (and use it) in the population from which the individual is drawn. As a partial remedy, we employed an internal-external validation as suggested by \citet{Steyerberg_Harrell_2016}. Internal-external validation is close to a V-fold CV, but the splitting is non-random. If the data are clustered according to some characteristics, such as centres; each centre becomes a fold. Then, we develop the prediction model on all centres but one and validate it on the remaining centre. Finally, we average the performance over the validation folds. We used this approach by clustering by individual time series, using a forward CV for each cluster, and averaging over the clusters. Thereby, POSL can also be viewed as a dynamic refitting process or model revision \citep{Jenkins_2018}. Finally, random-effects meta-analysis models could be used for verifying the homogeneity of the POSL's performance across individuals \citep{Austin_2016}. 

Sometimes, one might want to predict the convection volume \textit{conditionally on a particular intervention}, like using a wider needle to improve the blood flow rate \citep{Chapdelaine_2015}, because such predictions could be more informative. These \textit{counterfactual} prediction models are based on the same identification assumptions as for causal inference (see \citet{Chatton_Rohrer_2024} for an introduction) and need specific performance measures given the outcome is counterfactual (\textit{i.e.}, unobserved). We refer the interested readers to \citet{Efthimiou_2023} and \citet{Maas_2023} for more information on this topic.

\section{Conclusion}\label{sec7}

We demonstrated the potential of POSL to dynamically predict a repeated continuous outcome while optimising the prediction for its subject. We illustrated its performances on a real-life example in nephrology. However, this methodology can also be employed outside medical fields as long as sufficient observations are collected for the statistical unit. We then discussed several challenges involved in its use and our choices. We hope this study will encourage researchers to use this approach more broadly and provide guidance on using ML-based approaches. Furthermore, prediction being a necessary step for causality \citep{Hernan_Hsu_Healy_2019, vdLaan_Rose_2011}, POSL presents a new tool for personalised treatment rules \citep{Moodie_Stephens_2018}. Finally, prediction modellers should look beyond the statistical performances and technical challenges by considering the ethical issues of using ML \citep{Wiens_2019, Vollmer_2020}.

\section*{Competing interests}
No competing interest is declared.

\section*{Author contributions statement}

Conceptualisation: AC, MB, and MES; Data Curation: MB and RL; Formal Analysis \& Investigation: AC; Funding Acquisition: AC, RWP, and MES; Methodology: AC, RWP, and MES; Project Administration \& Resources: MES; Software \& Visualisation: AC; Supervision: RWP and MES; Writing - Original Draft: AC; and Writing - Review \& Editing: All authors.

\section*{Data Availability Statement}

Data may be available upon request from MB (email: michele.bally.chum@ssss.gouv.qc.ca) conditional on approval from the ethics committee of the Centre Hospitalier de l'Universit\'e de Montr\'eal. 
R codes are publicly available at https://github.com/ArthurChatton/Personalised-dynamic-SL.

\section*{Funding}
AC was supported by an IVADO postdoctoral fellowship (\#2022-7820036733). 

\section*{Acknowledgement}
 This project was approved by the Universit\'e de Montr\'eal ethic committee (\#2023-4344). A preprint of this manuscript is available here: 
https://doi.org/10.48550/arXiv.2310.08479.


\bibliographystyle{abbrvnat}
\bibliography{reference}

\section*{List of Figures}

\begin{enumerate}
    \item Rolling cross-validation schemes across four folds $v$.
    \item Flow-chart of the tuning-training-validation process.
    \item POSL’s forward cross-validation.
    \item Candidate learner distributions in the non-convex eSL and the dSL. (A) Percentage non-null weights in the eSL, (B) Sum of weights across time and individuals, (C) Percentage selection by dSL, (D) Number of learners composing the non-convex eSL over time, and (E) Sum of weights over time averaged between individuals.
    \item Time-pooled performance measures of the different POSL implementations and their candidate learners. (A) Accuracy assessed by the MdAE, (B) Accuracy assessed by the MSE, (C) Calibration-in-the-large (or mean calibration) assessed by the calibration intercept, and (D) Weak calibration assessed by the calibration slope. The dashed line represents the median of the best-performing method for accuracy measures or the ideal value for calibration measures.
    \item Performance measures of the different POSL implementations, the individual linear model, and the historical XGBoost. (A) Accuracy assessed by the MdAE, (B) Accuracy assessed by the MSE, (C) Calibration-in-the-large (or mean calibration) assessed by the calibration intercept, and (D) Weak calibration assessed by the calibration slope. (E) Calibration-in-the-small (or moderate calibration) of the POSLs at five time points. 
\end{enumerate}

\end{document}